\documentclass[journal]{IEEEtran}

\usepackage[pdftex]{graphicx}
\graphicspath{{./figures/}{./photos/}}
\DeclareGraphicsExtensions{.pdf,.jpeg,.png,.jpg}

\usepackage{amsmath, amsfonts, amssymb}
\usepackage{mathtools}
\usepackage[caption=false,font=small]{subfig}
\usepackage{stfloats}
\usepackage{url}
\usepackage{bm}			
\usepackage[per-mode=symbol]{siunitx}    		
\usepackage{cite}
\usepackage{tabularx}
\usepackage{booktabs}
\usepackage[shortlabels]{enumitem}
\usepackage{balance}

\DeclareSIUnit{\dBm}{dBm}	

\usepackage[xindy, shortcuts, acronym]{glossaries-extra}
\setabbreviationstyle[acronym]{long-short}
\glsdisablehyper
\newacronym{2d}{2D}{two-dimensional}
\newacronym{3d}{3D}{three-dimensional}
\newacronym{4g}{4G}{fourth generation}
\newacronym{5g}{5G}{fifth generation}
\newacronym{6g}{6G}{sixth generation}
\newacronym{adc}{ADCs}{analog-to-digital converter}
\newacronym{adcs}{ADCs}{analog-to-digital converters}
\newacronym{aoa}{AOA}{angle-of-arrival}
\newacronym{aod}{AOD}{angle-of-departure}
\newacronym{arfh}{ARFH}{auxiliary RF harvester}
\newacronym{asic}{ASIC}{application specific integrated circuit}
\newacronym{awgn}{AWGN}{additive white Gaussian noise}

\newacronym{bf}{BF}{beamformer}
\newacronym{bibc}{BiBC}{bistatic backscatter communication}
\newacronym{bs}{BS}{base station}
\newacronym{ccs}{CCS}{correlative channel sounder}
\newacronym{cdf}{CDF}{cumulative distribution function}
\newacronym{ce}{CE}{channel estimation}
\newacronym{ced}{CED}{cumulative energy demand}
\newacronym{csi}{CSI}{channel state information}
\newacronym{csp}{CSP}{Contact Service Point}
\newacronym{crlb}{CRLB}{Cram\'er-Rao lower bound}
\newacronym{da}{DA}{data association}
\newacronym{dc}{DC}{direct current}
\newacronym{dm}{DM}{diffuse multipath}
\newacronym{dmc}{DMC}{\gls{dm} component}
\newacronym{ecdf}{ECDF}{empirical cumulative distribution function}
\newacronym{ecsp}{ECSP}{Edge Computing Service Point}
\newacronym{eirp}{EIRP}{equivalent isotropically radiated power}
\newacronym{en}{EN}{energy neutral}
\newacronym{end}{END}{energy neutral device}
\newacronym{epu}{EPU}{edge processing unit}
\newacronym{fa}{FA}{federation anchor}
\newacronym{gbsc}{GBSC}{geometry-based deterministic stochastic channel}
\newacronym{gsm}{GSM}{Global System for Mobile Communications}  %
\newacronym{ic}{IC}{integrated circuit}
\newacronym{id}{ID}{information decoding}
\newacronym{imu}{IMU}{inertial measurement unit}
\newacronym{iot}{IoT}{Internet of Things}
\newacronym{lis}{LIS}{large intelligent surface}
\newacronym{los}{LoS}{line-of-sight}
\newacronym{lsa}{LSA}{large synthetic array}
\newacronym{lti}{LTI}{linear time-invariant}
\newacronym{mmwave}{mmWave}{millimeter wave}
\newacronym{mimo}{MIMO}{multiple-input multiple-output}
\newacronym{miso}{MISO}{multiple-input single-output}
\newacronym{mpc}{MPC}{multipath component}
\newacronym{mrt}{MRT}{maximum ratio transmission}
\newacronym{nfc}{NFC}{near-field communication}
\newacronym{ofdm}{OFDM}{orthogonal frequency-division multiplexing}
\newacronym{ota}{OTA}{over-the-air}
\newacronym{nb}{NB}{narrowband}
\newacronym{nlos}{NLoS}{non-line-of-sight}
\newacronym{pa}{PA}{power amplifier}
\newacronym{pl}{PL}{path loss}
\newacronym{pc}{PC}{pilot count}
\newacronym{rcs}{RCS}{radar cross section}
\newacronym{rf}{RF}{radio frequency}
\newacronym{rfid}{RFID}{radio frequency identification}
\newacronym{ris}{RIS}{reflective intelligent surface}
\newacronym{rms}{RMS}{root mean square}
\newacronym{rss}{RSS}{received signal strength}
\newacronym{rssi}{RSSI}{received signal strength indicator}
\newacronym{ra}{RA}{reflectarray}
\newacronym{rw}{RW}{RadioWeaves}
\newacronym{sa}{SA}{synchronization anchor}
\newacronym{sar}{SAR}{specific absorption rate}
\newacronym{sbl}{SBL}{sparse Bayesian learning}
\newacronym{sdn}{SDN}{software-defined network}
\newacronym{sinr}{SINR}{signal-to-interference-plus-noise ratio}
\newacronym{simo}{SIMO}{single-input multiple-output}
\newacronym{siso}{SISO}{single-input single-output}
\newacronym{slam}{SLAM}{simultaneous localization and mapping}
\newacronym{snr}{SNR}{signal-to-noise ratio}
\newacronym{smc}{SMC}{specular multipath component}
\newacronym{s-parameters}{S-parameters}{scattering parameters}
\newacronym{tosm}{TOSM}{through-open-short-match}
\newacronym{tdoa}{TDOA}{time-difference-of-arrival}
\newacronym{toa}{TOA}{time-of-arrival}
\newacronym{ue}{UE}{user equipment}
\newacronym{uhf}{UHF}{ultra high frequency}
\newacronym{ula}{ULA}{uniform linear array}
\newacronym{upa}{UPA}{uniform planar array}
\newacronym{ura}{URA}{uniform rectangular array}
\newacronym{uwb}{UWB}{ultrawideband}
\newacronym{vna}{VNA}{vector network analyzer}
\newacronym{wb}{WB}{wideband}
\newacronym{wpt}{WPT}{wireless power transfer}
\newacronym{zf}{ZF}{zero forcing}

\newacronym{ssf}{SSF}{small-scale fading}
\newacronym{lsf}{LSF}{large-scale fading}
\newacronym{ple}{PLE}{pathloss exponent}
\newacronym{usrp}{USRP}{universal software radio peripheral}
\newacronym{rb}{Rb}{rubidium}
\newacronym{pps}{1PPS}{1 pulse-per-second}
\newacronym{mrc}{MRC}{maximum ratio combinig}
\newacronym{fpga}{FPGA}{field-programmable gate array}
\newacronym{tdma}{TDMA}{time-division multiple access}
\newacronym{cp}{CP}{cyclic prefix}
\newacronym{dac}{DAC}{digital-to-analog converter}

\newacronym{pdf}{PDF}{probability density function}
\newacronym{cbsm}{CBSM}{correlation-based stochastic model}
\newacronym{gbsm}{GBSM}{geometry-based stochastic model}
\newacronym{vr}{VR}{visibility region}
\newacronym{acf}{ACF}{autocorrelation function}

\newacronym{jcas}{JCAS}{joint communication and sensing}
\newacronym{v2x}{V2x}{vehicle-to-everything}
\newacronym{m2m}{M2M}{machine-type communication}
\newacronym{dmimo}{D-MIMO}{distributed \glsentrylong{mimo}}
\newacronym{urllc}{URLLC}{ultra-reliable and low-latency communication}

\newacronym{cir}{CIR}{channel impulse response}

\newacronym{dli}{DLI}{direct link interference}
\newacronym{dpe}{DPE}{direct position estimation}
\newacronym{eh}{EH}{energy harvesting}

\newacronym{elaa}{ELAA}{extremely large aperture array}
\newacronym{em}{EM}{electromagnetic}

\newacronym{e2e}{E2E}{End-to-End}

\newacronym{esd}{ESD}{electrostatic discharge}

\newacronym{jcrs}{JCRS}{joint communication and radar sensing }

\newacronym{isac}{ISAC}{integrated sensing and communication}
\newacronym{ism}{ISM}{industrial, scientific and medical}
\newacronym{kpi}{KPI}{key performance indicator}
\newacronym{lca}{LCA}{life cycle assessment}

\newacronym{liion}{Li-ion}{lithium-ion}

\newacronym{lmo}{LMO}{lithium ion manganese oxide}
\newacronym{mc}{MC}{Monte Carlo}
\newacronym{mcu}{MCU}{microcontroller unit}
\newacronym{mf}{MF}{matched filter}
\newacronym{ml}{ML}{maximum likelihood}
\newacronym{mse}{MSE}{mean square error}
\newacronym{mmse}{MMSE}{minimum mean square error}

\newacronym{mobc}{MoBC}{monostatic backscatter communication}

\newacronym{mmimo}{mMIMO}{massive MIMO}

\newacronym{xlmimo}{XL-MIMO}{extremely large-scale MIMO}

\newacronym{mrfh}{MRFH}{main RF harvester}

\newacronym{olos}{OLoS}{obstructed LoS}

\newacronym{nr}{NR}{New Radio}

\newacronym{pf}{PF}{particle filter}

\newacronym{pdp}{PDP}{power delay profile}
\newacronym{dsd}{DSD}{Doppler spectral density}
\newacronym{per}{PER}{packet error rate}

\newacronym{pg}{PG}{path gain}

\newacronym{pw}{PW}{planar wavefront}

\newacronym{re}{RE}{radio element}

\newacronym{rmse}{RMSE}{root mean square error}

\newacronym{sdg}{SDG}{Sustainable Development Goal}

\newacronym{si}{SI}{self-interference}

\newacronym{s-parameter}{S-parameter}{scattering parameter}
\newacronym{svd}{SVD}{singular value decomposition}
\newacronym{sw}{SW}{spherical wavefront}
\newacronym{swipt}{SWIPT}{simultaneous wireless information and power transfer}

\newacronym{tdd}{TDD}{time division duplexing}

\newacronym{wsn}{WSN}{wireless sensor network}
\newacronym{xets}{XETS}{cross exponentially tapered slot}

\newacronym{ap}{AP}{access point}
\newacronym{fim}{FIM}{Fisher information matrix}
\newacronym{pcrlb}{PCRLB}{posterior Cram\'er-Rao lower bound}

\newacronym{eadf}{EADF}{effective aperture distribution function}
\newacronym{lhf}{LHF}{likelihood function}
\newacronym{wssus}{WSSUS}{wide-sense stationary and uncorrelated scatterers}
\newacronym{agc}{AGC}{automatic gain control}
\newacronym{ctf}{CTF}{channel transfer function}
\newacronym{lidar}{lidar}{light detection and ranging}
\newacronym{pll}{PLL}{phase-locked loop}
\newacronym{dpss}{DPSS}{discrete prolate spheroidal sequence}
\newacronym{icas}{ICAS}{integrated communication and sensing}
\newacronym{lo}{LO}{local oscillator}
\newacronym{grf}{GRF}{Gaussian random field}
\glsaddall

\newcommand{\complexset}[2]{ \mathbb{C}^{#1 \times #2}  } 
\newcommand{\realset}[2]{ \mathbb{R}^{#1 \times #2}  } 

\DeclareMathOperator{\vectorize}{vec}

\newcommand{\herm}{\mathsf{H}}
\newcommand{\trp}{\mathsf{T}}

\newcommand{\Rice}[2]{\operatorname{Rice} \left( #1, #2 \right) }

\newcommand{\Nfrequency}[0]{N_\text{\tiny f}}                       
\newcommand{\Ntime}[0]{N_\text{\tiny t}}                            

\newcommand{\fcarrier}[0]{\mathsf{f}_{\scriptscriptstyle \mathsf{c}}}                         

\newcommand{\channelVecNoisy}[2]{\widetilde{\bm{h}}_{\scriptscriptstyle#2}^{\scriptscriptstyle(#1)}}   

\newcommand{\channelMatrix}{\bm{H}^{(m)}}
\newcommand{\channelMatrixLOS}{\bm{H}_{\text{\tiny LoS}}^{(m)}}
\newcommand{\channelMatrixOLOS}{\bm{H}_{\text{\tiny OLoS}}^{(m)}}

\newcommand{\windowedH}{\bm{W}^{(w; m)}}
\newcommand{\localScatteringMatrix}[1]{\bm{C}^{(m)}_{#1}}

\begin{document}

\title{A Measurement-Based Spatially Consistent Channel Model for Distributed MIMO in Industrial Environments}

\author{Christian~Nelson,
        Sara~Willhammar,~\IEEEmembership{Member,~IEEE,}
        and~Fredrik~Tufvesson,~\IEEEmembership{Fellow,~IEEE}.
\thanks{Christian Nelson, Sara Willhammar and Fredrik Tufvesson are with the Department of Electrical and Information Technology, Lund University, Lund,
Sweden. Email: {firstname.lastname}@eit.lth.se}
\thanks{This work has been supported by the European Union’s Horizon 2020 research and innovation program under grant agreement No 101013425 (REINDEER), by the Excellence Center at Linköping – Lund in Information Technology (ELLIIT) and the Vinnova competence center NextG2com.}%
\thanks{Manuscript received Dec. 17, 2024; revised March 31, 2026, revised July 4, 2026.}}

\markboth{IEEE Transactions on Wireless Communications,~Vol.~XX, No.~YY, Date}%
{Shell \MakeLowercase{\textit{et al.}}: Bare Demo of IEEEtran.cls for IEEE Journals}
%

\maketitle

\begin{abstract}
Future wireless communication systems are envisioned to support ultra-reliable and low-latency communication (URLLC), which will enable new applications such as compute offloading, wireless real-time control, and reliable monitoring. Distributed multiple-input multiple-output (D-MIMO) is one of the most promising technologies for delivering URLLC. This paper classifies obstructions and derives a channel model from a D-MIMO measurement campaign carried out at a carrier frequency of 3.75\,GHz with a bandwidth of 35\,MHz using twelve fully coherent distributed dipole antennas in an industrial environment. Channel characteristics are investigated, including statistical measures such as small-scale fading, large-scale fading, delay spread, and transition rates between line-of-sight and obstructed line-of-sight conditions for the different antenna elements, providing the foundation for an accurate channel model for D-MIMO systems in industrial environments. Furthermore, to ensure spatially consistent simulation results, the correlations of large-scale fading between antennas are modeled using Gaussian random fields. Lastly, the tails of the fitted distributions are modeled to enable accurate evaluation of reliability and rare events. Based on the results, a channel model for D-MIMO in industrial environments is presented together with a recipe for its implementation and a validation.
\end{abstract}


\IEEEpeerreviewmaketitle

\section{Introduction}

\IEEEPARstart{F}{uture} wireless systems are required to support the ultra-reliable low-latency communication (URLLC) use case for applications such as remote driving and industrial automation. In addition, the latter also needs functionalities such as localization, sensing, and the ability to support energy neutral devices, to name a few~\cite{itur6g}.
As part of enabling these functionalities and meeting the use case and application requirements, there is a technological shift toward distributed computing capabilities and \gls{dmimo}\glsunset{mimo} systems~\cite{VanderPerre2019}. 

The shift to more \gls{dmimo} systems has a large potential to increase the reliability. Increasing the number of antennas in a \gls{mimo} system has been shown to improve reliability by decreasing small-scale fading, through the so-called channel hardening effect~\cite{Willhammar2020}. \Gls{dmimo} systems have the potential to further improve the system stability, as they will also reduce large-scale fading~\cite{Flordelis2015, Willhammar2024}. This is because there will be -- with high probability -- one or several access points close by, experiencing different large-scale fading effects. Decreasing the fading effects not only improves the reliability but also reduces the required transmit power and needed fading margins. 
\Gls{dmimo} systems have the potential to decrease both of these fading effects, and due to the distributed setup, correlations of large-scale fading are of special interest. These correlations are  essential channel characteristics that need to be captured in channel models that can be used for accurate and realistic simulations of new systems, use cases, and functionalities, as well as for product development.

For such simulations, a fully deterministic ray tracer provides accurate results, but the results would be restricted to a specific environment. Adding stochastic elements to a ray-tracer would improve generality while preserving spatial consistency, enabling the evaluation of sensing algorithms. Conversely, a fully stochastic channel model based solely on extracted channel parameters can reproduce channel statistics, but may struggle to preserve physical consistency for sensing, localization, and non-stationary propagation effects.

To ensure that the essential characteristics of the channel for the target scenario are captured in the model, measurement data are required. Wireless channels in industrial environments have been measured and modeled for decades \cite{Rappaport1991, hampdicke1999, Karedal2004, Holfeld2016, Wassie2019, Zhang2022b, Willhammar2024}. 
However, as system architectures evolve, there is a need for new channel models to capture channel propagation characteristics that previously were either unattainable or of no interest given the system under consideration.
In \cite{photonics12030257} the authors summarize the efforts of investigating an industrial environment and then present a model from sub-6\,GHz up to visible light channels. Their results are based on ray-tracing simulations and not measurements and only cover static channel parameters.
In \cite{measIOT2020} the two bands \SI{3}{\giga\hertz} to \SI{4}{\giga\hertz} and \SI{38}{\giga\hertz} to \SI{40}{\giga\hertz} are measured and evaluated in an industrial environment, but no dynamic or spatial correlation is investigated and they do not conduct any investigation of distributed \gls{mimo}.
The authors of \cite{propcharactIndustrialEnnv2021} conduct a measurement campaign at \SI{4.1}{\giga\hertz} in a single-link static configuration.
A directional investigation to industrial areas is also carried out in \cite{Sub6ChannelModel2022} but in a static industrial environment.
In the 3GPP technical report 38.901 Rel.~18~\cite{3gppCHANNELMODEL}, general propagation channel models are presented for indoor factories. Specifically the scenario called ``Indoor Factory with Dense clutter and High base station height (InF-DH)'' is the most relevant here.
An industrial environment most likely consists of large obstructing objects in the form of shelves or large machinery. The industrial environment is often cluttered with these types of obstructions. An indoor office, on the other hand, can be long hallways with single room offices along the way, or an open landscape but with much less obstruction.
In~\cite{Guevara2021, Choi2020}, measurement-based comparisons were made between different antenna configurations.
In \cite{Guevara2021}, the scenario is static and only \gls{rss} is presented to evaluate the coverage. The authors also investigate how co-located and distributed configurations compare when a maximum ratio precoder is used to beamform to a user, and how much power that is leaked to other idle users. \cite{Choi2020} performed a similar analysis but instead focused on the achievable downlink spectral efficiency. In another static indoor distributed setup, a measurement-based estimation of the large-scale parameters and its spatial consistency was carried out and the path gain and the large-scale fading were simultaneously modeled using a Gaussian process~\cite{Jang2022}. There, small-scale fading was not considered.
Common to these works is that they consider static or single-link settings, or rely on ray-tracing rather than measurements, or do not address the joint spatial correlation of the large-scale fading across distributed antennas. It is this combined gap that motivates the measurement-based, spatially consistent D-MIMO model developed here.

In this work, we have conducted a unique measurement campaign with a  truly coherent and parallel D-MIMO system in an industrial environment. 
We extend the work in \cite{Nelson2024} by performing a thorough statistical analysis of the measurements.
We present a new approach for classifying obstruction in D-MIMO systems. We investigate spatial and temporal non-stationarities and extract key channel parameters, including fading characteristics and joint large-scale fading correlations, to develop an accurate yet simple measurement-based spatially consistent channel model. This model can be used for realistic system simulations of D-MIMO systems in industrial scenarios. 
The main contributions of this paper are as follows. First, we perform a statistical analysis of truly coherent, parallel dynamic D-MIMO measurements in an industrial environment. 
Second, we analyze stationarity regions, fading parameters and joint correlation of the large-scale fading. 
Finally, based on the results, we present a novel comprehensive spatially consistent channel model showing good agreement with the measurements. 
We also provide the data set and an open source implementation of the proposed channel model.

The paper is structured as follows. First, in Section~\ref{sec:measScene}, the measurement equipment and environment are introduced. Then in Section~\ref{sec:classifying-obstruction}, we classify obstruction as caused by the environment, classifying samples as \gls{los} and \gls{olos}. From that we compute the path gain in Section~\ref{sec:pathGain} and the large-scale fading parameters in Section~\ref{sec:lsp}. This is followed by Section~\ref{sec:ssf}, which includes an analysis of the small-scale fading statistics, where the local scattering function and collinearity are computed to get to the appropriate stationarity regions. A channel model recipe to generate spatially consistent D-MIMO channels in industrial environments is presented in Section~\ref{sec:channelModel} and validated in Section~\ref{sec:validation}. Finally, conclusions are given in Section~\ref{sec:conclusions}.

{\slshape Notation:}
Column vectors and matrices are denoted by boldface lowercase, $\bm{x}$, and uppercase letters, $\bm{X}$, respectively.
The operator $\vectorize: 
\bm{X}\mapsto \bm{x}$ vectorizes the matrix $\bm{X}$
by stacking its columns on top of each other in $\bm{x}$.
$\left(\cdot\right)^\trp$ and $\left(\cdot\right)^\herm$ denote the transpose and the Hermitian transpose of $\left(\cdot\right)$, respectively.
The element-wise multiplication (Hadamard product) is denoted by $\odot$ and the convolution is denoted by~$*$.
The Euclidean norm of the vector $\bm{x}$ is $\lVert \bm{x} \rVert$. 
The $N\times N$ identity matrix is denoted as $\bm{I}$, where the size is implicit from the context.
Further, $\bm{1}$ is a column vector where all elements are~1 and its length is implicit from context. $\overline{\bm{X}}$ means the element-wise complex conjugate of $\bm{X}$. 
Lastly, $\left[\bm{X}\right]_{m,n}$ is the element in the $m$-th row and $n$-th column.
\begin{figure}[t]
    \centering
    \includegraphics[width=0.99\columnwidth, trim={0 40pt 0 20pt}, clip]{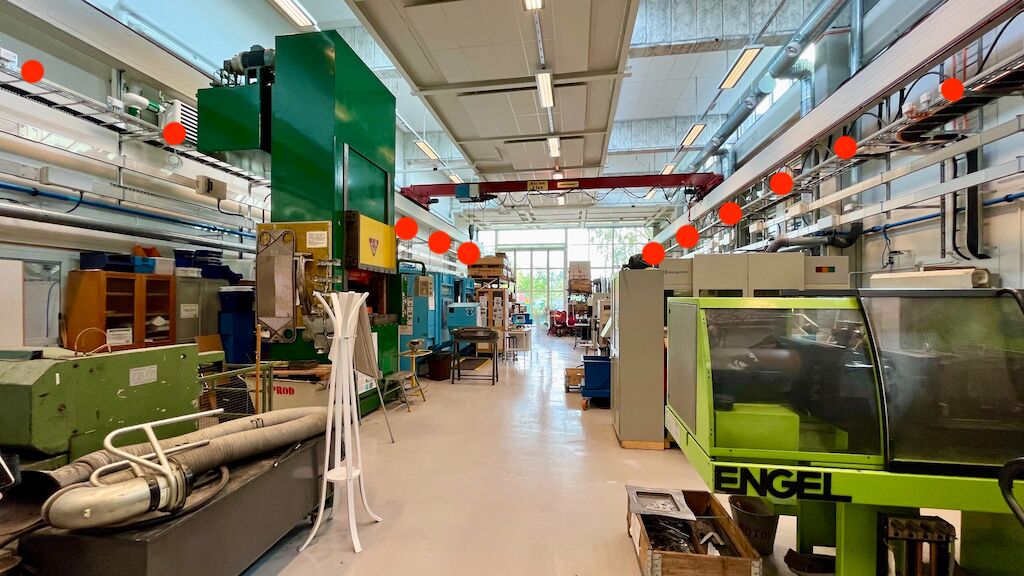}
    \caption{The rich scattering and heavily shadowed industrial environment where the D-\gls{mimo} measurements were conducted. The dots mark the approximate locations of the antennas.}
    \label{fig:photo-hall}
\end{figure}

\section{Measurement scenario}\label{sec:measScene}
The measurement campaign was carried out in an industrial environment in the department of mechanical engineering at Lund University with details presented in \cite{Nelson2024}. 
The measurement environment is depicted in Fig.~\ref{fig:photo-hall} where some of the machinery can be seen. Most of the machinery was enclosed and the radio environment was fairly static, except for the sporadic movement of people working in the hall.
The channel sounder presented in \cite{Nelson2024} was used to collect the samples at a carrier frequency of $\fcarrier = \SI{3.75}{\giga\hertz}$, utilizing a bandwidth of \SI{35}{\mega\hertz} distributed over $\Nfrequency = 449$ tones, resulting in a carrier spacing of \SI{78.125}{\kilo\hertz}.
The sounder is designed for D-\gls{mimo} measurements and captures all possible link combinations that later can be used for offline processing and analysis.
All included radio units share the same notion of time through a distributed \gls{pps} signal, 
and are frequency disciplined with external coherent \gls{rb} clocks.

\begin{figure}[t]
    \centering
    \includegraphics{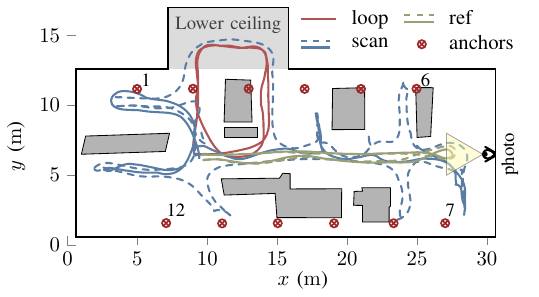}
    \caption{Top-down overview of the industrial environment where the D-\gls{mimo} measurements were conducted. The photograph in Fig.~\ref{fig:photo-hall} is taken from the position indicated by the eye. The anchors are visible along the two sides and the trajectories driven by the agent are visualized. The numbering of the anchors is shown for anchors 1, 6, 7, and~12; the rest are implicit for visual purposes. The solid and dashed lines represent different measurement runs, i.e. measurements captured during different times.}
    \label{fig:scenarios}
\end{figure}

Fig.~\ref{fig:photo-hall} depicts the measurement environment, which is characterized as a rich scattering environment with many metallic objects and structures, causing both reflections and shadowing. As shown in Fig.~\ref{fig:scenarios}, the dimensions of the room are approximately \SI{30}{\metre} long and \SI{12}{\metre} wide, with a ceiling height of \SI{8}{\metre}. The access points, here called anchors, were mounted on rails approximately \SI{4}{\metre} above the floor with antennas tilted downward to cover the center of the room.

At each anchor, the output of the sounder is a channel vector $\channelVecNoisy{m}{k} \in \complexset{\Nfrequency}{1}$, at time $k$ and from anchor $m$ for the $\Nfrequency$ tones, which is an estimate of the true channel transfer function, but corrupted by noise. 
For this campaign, $m \in \{1, 2, ..., 12\}$ and $k \in \{0, 1, ..., \Ntime -1\}$, where $\Ntime$ is the number of measured snapshots. 
All captured channel transfer functions from a measurement are collected in the matrix
\begin{equation}
    \channelMatrix = \left[\channelVecNoisy{m}{0}, \ldots, \channelVecNoisy{m}{\Ntime - 1}\right] \in \complexset{\Nfrequency}{\Ntime}
\end{equation}
for further processing.
The agent moving in the environment was a remotely controlled robot that carried all the necessary radio equipment.
To obtain an estimate of the ground truth position of the agent on the map during the measurements, the agent was equipped with a \gls{lidar} and an \gls{imu} to perform offline \gls{slam}~\cite{Xu2022}.
A more detailed list and description of the equipment can be found in \cite{Nelson2024}.

Here, five measurements are presented covering three different scenarios, as visualized in Fig.~\ref{fig:scenarios}. 
The three scenarios are: 1) the ref scenario where the agent was driving back and forth in the middle of the hall at a speed of approximately \SI{0.8}{\metre\per\second}, 2) the loop scenario where the agent was driving two laps around machinery (parts of the loop are in a section of the industrial hall where the ceiling is considerably lower, which will lead to a challenging radio channel environment with a lot of obstruction), and 3) the scan scenario where the agent was driving around to cover most of the accessible area of the hall, including in between machines. 
With \num{449} active tones, for \num{5}~measurements, each with a length between \SI{60}{\second} and \SI{240}{\second}, the result is \num{1440000} recorded channel transfer functions.
The raw data are available at~\cite{zenodo}.

\begin{figure}[t]
    \centering
    \includegraphics{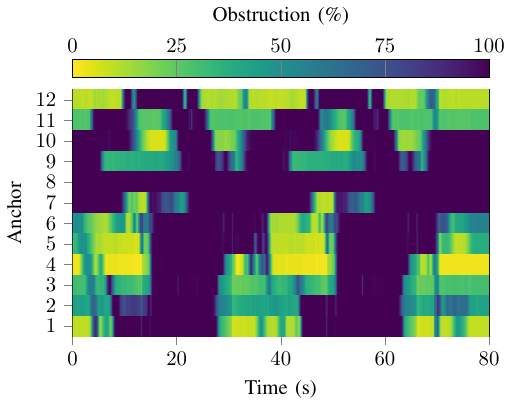}
    \caption{The amount of obstruction of the (approximated) first Fresnel ellipse, over time for all anchors. The scenario depicted is the loop scenario.}
    \label{fig:raw-data}
\end{figure}

\section{Classifying obstruction}\label{sec:classifying-obstruction}
\begin{figure}[t]
\centering
\subfloat{\includegraphics{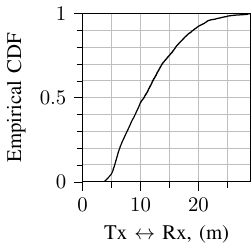}%
\label{fig:distances-cdf}}
\hfil
\subfloat{\includegraphics{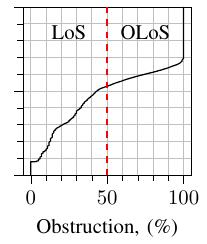}%
\label{fig:fresnel-cdf}}
\caption{Left: The distribution of the anchor and agent separation. Right: the \gls{ecdf} of the approximated obstruction of the first Fresnel ellipse. From the \gls{ecdf} of the obstruction it is clear that the environment in which the data has been collected is challenging, with a lot of obstructing machinery between the agent and the anchors. The \gls{ecdf} shows a 100\,\% obstruction in more than 30\,\% of the collected data, given the choice of a $2\lambda$ radii of the cylinder.}
\label{fig:scenario-data}
\end{figure}

In D-MIMO systems, it is likely that one or more anchors are in \gls{los} while others are in \gls{olos}. To quantify the probability of being in either state, the obstruction is classified in the following section. An approximation of the obstruction of the first Fresnel zone can be derived from the \gls{lidar} data and used for the purpose of state classification.
As a result of this, the obstruction of the first Fresnel zone is seen in Fig.~\ref{fig:raw-data} for the loop scenario in Fig.~\ref{fig:scenarios}. There are variations over time, where the link between the agent and an anchor is more or less obstructed. At each instant of time, at least some anchors have a more or less \gls{los} link, i.e. a low amount of obstruction of the Fresnel ellipse, while some are more obstructed. As an example of the latter, anchor~8 is always obstructed with an obstruction of 100\%. It is also clear that when the agent is driving into the section with a lower ceiling, the \gls{los} component is completely obstructed to all the anchors on one side of the room.

The link obstruction was computed as follows.
The output from the \gls{slam} algorithm provides the ground truth trajectory of the agent and a \gls{3d} point cloud representation of the environment, i.e. a map. 
With these data, the obstruction of the first Fresnel zone can be approximated, as elaborated in the following. 
In each time step $k$, and for each anchor $m$, the line between the agent and the anchor $m$ is drawn. 
Then all \gls{3d} \gls{lidar} points that have a distance to the line that is less than a given radius $r$ are saved.
Those lidar points are then projected onto a circle in the cylinder. Choosing the radius $r$ related to the size of the first Fresnel zone enables the derivation of a metric of the fraction of the circle area that is covered with lidar points, i.e. objects in the environment. 
This is an approximation of the amount of obstruction between the anchor $m$ and the agent at time $k$, that is, the Fresnel obstruction. 
Let $\mathbf{LOS} \in \realset{M}{\Ntime}$ denote the obstruction of the circle at time $k$ for the link $m$ where $\left[\mathbf{LOS}\right]_{m,k} \in \left[0, 100\right]$, 
since the approximation of the first Fresnel zone depends on the cylinder radius, the radius of the Fresnel ellipse was determined for a distance of \SI{5}{\metre}. This choice was based on the fact that all machinery was located along the walls, close to the anchors, and provides a reasonable trade-off in the resulting obstruction estimates.
Furthermore, to be less sensitive to noise in the \gls{lidar} scans, a manual investigation of the radius was also conducted. Then the radius was chosen as $r = 2\lambda$.

The \gls{ecdf} for the distances between agent (Tx) and anchor (Rx) is shown in Fig.~\ref{fig:distances-cdf} and the \gls{ecdf} of the approximated obstruction of the first Fresnel zone is shown in Fig.~\ref{fig:fresnel-cdf}. 
Due to the environment and deployment, it is rare that the distance between the agent and an anchor is above 20~m, as depicted in Fig.~\ref{fig:distances-cdf}.
In the \gls{ecdf} in Fig.~\ref{fig:fresnel-cdf} it can be seen that this is a challenging environment where the links in the dataset are in a fully obstructed state more than \SI{30}{\percent} of the time.

The data is classified into two states. One in which obstruction lies between \SIrange{0}{50}{\percent}, and another in which the obstruction is in the range \SIrange{50}{100}{\percent}. 
The states are named \gls{los} and \gls{olos}, respectively. The threshold was determined based on empirical analysis and means that if less than 50\% of the first Fresnel zone is obstructed, then it is classified as \gls{los}. 
The chosen threshold had little impact on further evaluations. 
A brief sensitivity analysis was made to evaluate the impact of different thresholds for the path gain and large-scale model parameters, concluding that the variation of the parameters was between 0.1 and 0.3~dB, except for the m-parameter in the path gain model where it in \gls{olos} was 3.2~dB. 
The 50\% threshold gives parameters in line within the mean obtained for the different thresholds within 0.2~dB, again with the largest discrepancy for the \gls{olos} m-parameter with a difference of 1.8~dB. 
For completeness, it should be mentioned that the slightly more complex case with three states \gls{los} (\SI{0}{\percent} obstruction), \gls{olos} (\SIrange{1}{99}{\percent} obstruction) and \gls{nlos} (\SI{100}{\percent} obstruction) was also analyzed, but showed no major differences in the statistics that would motivate this more complex classification and modeling approach. 
Hence, the matrix $\channelMatrix$ is split into matrices $\channelMatrixLOS$ and $\channelMatrixOLOS$, where the difference between two consecutive time indices no longer needs to be equal, i.e.,
\begin{equation}
     \channelMatrixLOS =\left[\channelVecNoisy{m}{k} | k \in \{ 0, \ldots, \Ntime - 1 \} : \left[\mathbf{LOS}\right]_{m,k} \leq 50 \right],
\end{equation}
and
\begin{equation}
     \channelMatrixOLOS =\left[\channelVecNoisy{m}{k} | k \in \{ 0, \ldots, \Ntime - 1 \} : \left[\mathbf{LOS}\right]_{m,k} \geq 50 \right].
\end{equation}

Having classified the obstruction and divided the dataset into \gls{los} and \gls{olos}, the next steps to create a channel model include computing the probability that there is a \gls{los} link and the probabilities of a change of state, i.e. the probabilities of going from \gls{los} to \gls{olos}, and vice versa. 

\subsection{Line-of-sight probability and state transitions}\label{sec:losProbability}

\begin{figure}[t]
	\centering
	\includegraphics{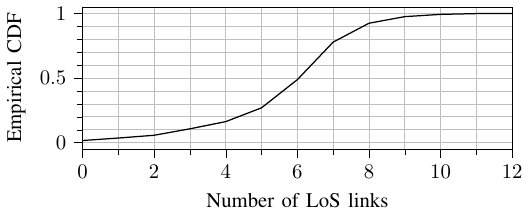}
	\caption{The \gls{ecdf} of the number of links  with \gls{los} conditions throughout the measurements. }
	\label{fig:ecdf-links}
\end{figure}

Starting from the two datasets as classified and divided in the previous section and to get a better understanding of what constitutes real D-MIMO channels in heavily shadowed industrial environments, the distribution of the number of \gls{los} links in each time instance is shown in Fig.~\ref{fig:ecdf-links}.

It can be observed that in this scenario there are never twelve links in \gls{los} condition at the same time. However, there is also a low probability that there are zero links in \gls{los}, which is promising from a reliability perspective. In these measurements, there are six antennas (50\% of the total number) with the \gls{los} condition, 50\% of the time. 
\begin{figure}[!ht]
    \centering
	\includegraphics{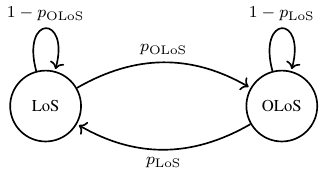}
    \caption{State transition graph, including the probability to stay in \gls{los} and \gls{olos} as well as to change to \gls{olos} and to \gls{los} from the respective state.}
    \label{fig:states}
\end{figure}
In the following, the average number of state transitions per distance traveled is estimated; that is, how often a link alternates between states \gls{los} and \gls{olos}. These two states are visualized in Fig.~\ref{fig:states}, where a transition between states occurs with a probability $p_s$; the probability of staying in a certain state is then $1-p_s$, for $s \in \{\mathrm{LoS},\mathrm{OLoS}\}$.
Each anchor $m$ has its own probabilities for the transitions. 
For reference, in this scenario, the average distance that an anchor is classified as \gls{los} is $\SI{4.65}{\metre}$ and for \gls{olos} the corresponding value is $\SI{5.55}{\metre}$.

\section{Path gain}\label{sec:pathGain}
\begin{figure}[!t]
\centering
\subfloat{\includegraphics[]{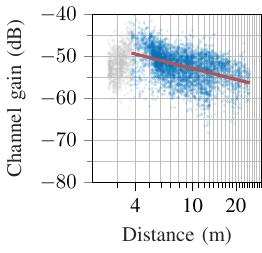}}%
\label{fig:path-gain-los}
\hfil
\subfloat{\includegraphics[]{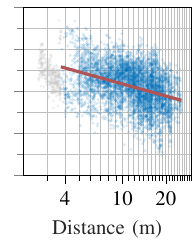}}%
\label{fig:path-gain-olos}
\caption{The channel power gains and the corresponding linear model for the \gls{los} (left) and \gls{olos} (right) datasets, respectively. The gray points are discarded measurements due to the large impact of the combined antenna gain patterns. Note that only every 150th point is shown in the plot for clarity.}
\label{fig:path-gain}
\end{figure}

Having the obstruction classification and the state transition probability in place, the next step is to model the distance-dependent path gain for the two states.
\begin{figure}[t]
    \centering
    \includegraphics{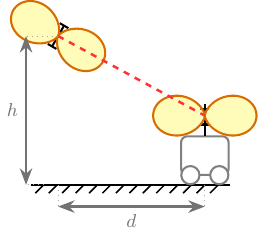}
    \caption{A schematic description of the influence of the antenna gain depending on the distance. As the agent distance d decreases the (theoretical) nulls in the antenna pattern start to have a big influence on the received signal power. This is clearly seen in Fig.~\ref{fig:path-gain}.}
    \label{fig:ant-schematic}
\end{figure}
Starting by averaging the channel power gains over the tones as in
\begin{equation}
\bar{\bm{P}}^{(m)} = \frac{1}{\Nfrequency} \bm{1}^\trp\left(\channelMatrix \odot \overline{\channelMatrix}\right) \in \realset{1}{\Ntime},
\end{equation}
and then collecting the averaged power gains from all anchors $m$ in the matrix
\begin{equation}\label{eq:avg-rx-pwr}
    \bar{\bm{P}} = \left[\bar{\bm{P}}^{(1)^\trp}, \ldots, \bar{\bm{P}}^{(M)^\trp}\right] \in \realset{\Ntime}{M},
\end{equation}
then the matrix 
\begin{equation}\label{eq:distance-matrix}
    \bm{D} = \left[ \bm{d}^{(1)}, \ldots, \bm{d}^{(M)} \right]\in \realset{\Ntime}{M} 
\end{equation}
contains the distances from the agent to all the anchors, where the distribution of the distances across all measurements is shown in Fig.~\ref{fig:distances-cdf}. Vectorizing the matrices
\begin{equation}
    \vectorize{\bar{\bm{P}}}\in \realset{M\Ntime}{1}\quad\text{and}\quad \vectorize{\bm{D}} \in \realset{M\Ntime}{1},
\end{equation}
and sorting them simultaneously leads to a monotonically increasing $\vectorize{\bm{D}}$. Since the channel coefficients in $\channelMatrix$ contain the influence from the antennas (and their antenna gain patterns), all points closer than $2.65\cdot\sqrt{2}$ meters are removed when estimating the path gain parameters. This distance corresponds to an elevation angle of 45 degrees. 
The vertical height difference between an anchor and the agent antenna is approximately \SI{2.65}{\metre}, meaning that a $45^{\circ}$ elevation corresponds to a horizontal separation of about \SI{2.65}{\metre} and hence a link distance of about $2.65\cdot\sqrt{2}$~m.
Below this angle, the antenna gain of the dipole antenna is more or less flat. But above it, the antenna gain drops and heavily influences the extracted path loss parameters, see Fig.~\ref{fig:ant-schematic}.
The log-distance results for both the \gls{los} and \gls{olos} state are shown in Fig.~\ref{fig:path-gain}, where the gray dots are the discarded measurement points. The minimum mean square error (MMSE) lines in red are computed as
\begin{equation}\label{eq:path-gain}
    \text{PG}\left(d\right) = 
    \begin{dcases*}
        -44.24 - 0.86 \cdot 10\log_{10}\left(d/d_0\right),\quad \text{LoS}, \\
        -48.78 - 0.95 \cdot 10\log_{10}\left(d/d_0\right),\quad \text{OLoS}.
    \end{dcases*}
\end{equation}
for all distances $d \in \left[2.65\cdot\sqrt{2}, \,\,30 \right]$ meters. The reference distance is set at $d_0 = 1$~m. As seen in Fig.~\ref{fig:path-gain}, the \gls{los} dataset has in general a higher channel gain and the measurement points show smaller variations around the MMSE line, in comparison to the \gls{olos} dataset.
The estimated path loss exponents are relatively small compared with values reported in previous measurement campaigns that have been performed over the years~\cite{Rappaport1989, Karedal2004}, which could be a result of the rich scattering and the relatively short link distances and that the antennas are considered a part of the channel in this dataset. The combined antenna gain patterns from the anchors and the agent also make the slope less steep.

\section{Large-scale fading}\label{sec:lsp}
\begin{figure}[t]
	\centering
    \includegraphics[]{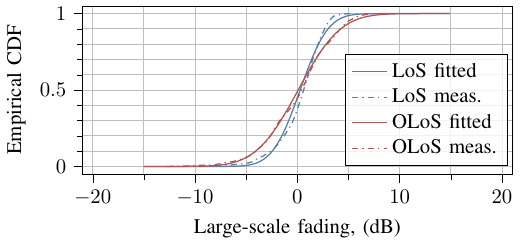}
	\caption{The \gls{ecdf} of the large-scale fading with the corresponding fitted log-normal distribution, for the \gls{los} and \gls{olos} dataset, respectively.}
	\label{fig:large-scale-cdf}
\end{figure}

To estimate large-scale fading, the model of the distance-dependent path gain in \eqref{eq:path-gain} is subtracted from the measured average power gain \eqref{eq:avg-rx-pwr}, i.e. $\bar{\bm{P}} - \mathbf{PG}\left(\bm{D}\right)$ where $\bm{D}$ is the measured distance in \eqref{eq:distance-matrix}. Then, a moving average window with a length of approximately $10\lambda$ is applied to estimate large-scale fading~\cite{Molisch2012}. The resulting distribution of the large-scale fading follows a log-normal distribution and the \gls{ecdf} of the estimated large-scale fading with the corresponding fitted distributions are shown in Fig.~\ref{fig:large-scale-cdf}, for both the \gls{los} and \gls{olos} datasets. The means for the two distributions are sufficiently close to zero that they are assumed to be zero with their respective variances. The mean and standard deviation for the \gls{los} dataset is \SI{0.27}{\dB} and \SI{2.13}{\dB}, respectively, and the corresponding values are \SI{0.08}{\dB} and \SI{3.25}{\dB} for the \gls{olos} dataset. As expected, the variation for the \gls{olos} case is slightly larger.

\subsection{Covariance}\label{sec:lsp-covariance}
\begin{figure*}[t]
\centering
\subfloat{\includegraphics{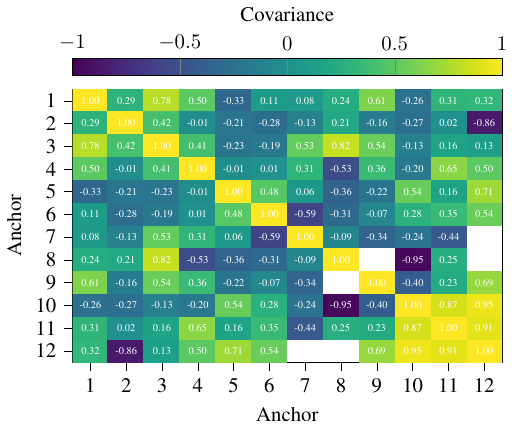}%
\label{fig:lsf-cov-los}}
\hfil
\subfloat{\includegraphics{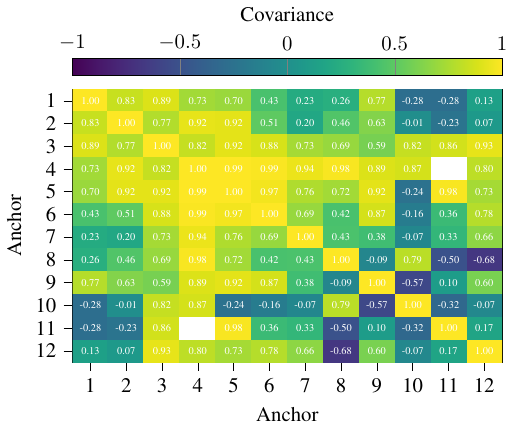}%
\label{fig:lsf-cov-olos}}
\caption{The covariance of the large-scale fading for \gls{los} (left) and \gls{olos} (right). Each entry in the matrix has been estimated pairwise to get enough data. The white entries indicate anchor pairs with insufficient simultaneous data to estimate the covariance: $(7,12)$, $(8,12)$, and $(8,9)$ in \gls{los}, and $(4,11)$ in \gls{olos}.}
\label{fig:lsf-cov}
\end{figure*}

For distributed systems, the correlation of large-scale parameters becomes important, as it will affect system performance~\cite{Dahman2014, Dahman2018}. 
The cosine similarity (reflective correlation) is used to compute the covariance of large-scale fading between the anchors, i.e. the mean is not removed. The sample reflective correlation between the two vectors $\bm{x}$ and $\bm{y}$ can be calculated as (in this case the vectors represent the large-scale fading time series for two anchors)
\begin{equation}
    \rho_{xy} = \frac{\sum x_k y_k}{\sqrt{\sum x_k^2}\sqrt{\sum y_k^2}}.
\end{equation}
The resulting covariance of large-scale fading between all anchors can be seen in Fig.~\ref{fig:lsf-cov} for the \gls{los} and the \gls{olos} datasets, respectively. Naturally, the diagonal in both cases is one. In the \gls{los} dataset, there are not enough simultaneous \gls{los} links to compute the correlation for the anchor pairs $(7,12)$, $(8,12)$, and $(8,9)$, while in the \gls{olos} dataset the pair $(4,11)$ lacks sufficient data; these missing pairs appear as the white entries in the covariance matrices. One general observation is that the covariances are smaller in the \gls{los} dataset than in the \gls{olos} dataset; the main exception being anchors 10, 11 and 12, which tend to have more similar statistics in \gls{los}.

For the \gls{olos} dataset, the covariance of the large-scale fading is more prominent. This can especially be observed between anchors situated on the same side of the wall and even more for anchors 1--6,
which seem to be interacting with the same objects at the same time, leading to them experiencing similar large-scale fading behavior. The cross-correlation between anchors on opposite sides tends to have smaller values; a tendency that can also be observed in the \gls{los} dataset, although not as prominent.

From the geometry of the anchors (Fig.~\ref{fig:scenarios}) it is clear that we can divide the covariance matrix into four sub-matrices: 1) anchors 1--6 on one side of the room with equidistant spacing of 4 meters, 2) anchors 7--12 on the other side of the room, lastly 3) and 4) the covariance terms that relate to anchors on opposite sides of the room.
Using the available data, the standard deviation of the off-diagonal elements in each sub-matrix was estimated as $\sigma \approx 0.28.$ This indicates that inter-anchor distance alone does not explain the covariance; environmental factors also have a substantial influence.
In particular, the same-side sub-matrices (anchors 1--6 and 7--12) contain anchor pairs whose separations range from \SI{4}{\metre} up to about \SI{20}{\metre}, yet the variance of the off-diagonal elements is similar across all four sub-matrices. If the inter-anchor distance were the dominant factor, the same-side and cross-side sub-matrices would be expected to differ markedly, so this similarity points to a substantial environmental contribution to the covariance.
Therefore, a more general approach was adopted in which each anchor covariance is drawn independently. Realizations of the truncated normal covariance distribution will therefore naturally produce groups of anchors exhibiting high covariance.
 The geometric structure visible in Fig.~\ref{fig:lsf-cov} is therefore not reproduced deterministically; instead, it is captured stochastically in the channel model later presented in Sec.~\ref{sec:channelModel}, where the measured covariance (Sec.~\ref{sec:lsp-covariance}) is combined with a distance-dependent kernel and an environment term to draw spatially consistent realizations.

\subsection{Auto-correlation}\label{sec:lspAutocorrelation}
Having evaluated the instantaneous covariance between anchors, in the following section, the auto-correlation $R_{xx}^{(m)}$ of the large-scale fading per anchor was explored. For the two datasets the auto-correlation was computed for all anchors and the decay factor $k$ was estimated in the following~\cite{Gudmundson1991}
\begin{equation}
    R_{xx}^{(m)}(d)=e^{-kd}.
\end{equation}
The resulting values for $k$ are: $0.82$ in the \gls{los} dataset and $0.81$ in the \gls{olos} dataset .
Defining the distance $d_{1/\mathrm{e}}$ such that the auto-correlation  $R_{xx}(d_{1/\mathrm{e}})=1/\mathrm{e}$ as the decorrelation distance yielding decorrelation distances of \SI{1.22}{\metre} and \SI{1.24}{\metre} for the \gls{los} and \gls{olos} datasets, respectively. The similarity between these two exponential decay rates could in part be explained by the splitting of the dataset into \gls{los} and \gls{olos}. Investigating Fig.~\ref{fig:fresnel-cdf} shows that the majority of the data are in an \gls{olos} state.

\section{Small-scale fading}\label{sec:ssf}
To characterize the small-scale fading statistics, one first has to define the regions where the channel can be considered as \gls{wssus}. This is done through the local scattering function and by analyzing the collinearity. 

\subsection{The Local Scattering Function}
The local scattering function is typically used to characterize dynamic (non-stationary) channels~\cite{Matz2005, BernadoPhD}. When moving in a rich scattering and heavily shadowed environment, such as the industrial hall presented here, one might violate the \gls{wssus} assumption conventionally used, if considering the entire dataset at once.
By treating the channel as a piece-wise stationary stochastic process -- i.e. the data is windowed into smaller parts where the \gls{wssus} assumptions approximately hold -- it is possible to extract the channel parameters in this local region.
The local scattering function solves the problem by applying a time- and frequency-bound filter that moves over the dataset. Following the methodology presented in \cite{BernadoPhD}, the sliding window indices in time and frequency are denoted as $k_t$ and $k_f$, respectively, and the sizes of the windows in time and frequency are denoted as $K$ and $N$, respectively.
Following \cite{Zelenbaba2021}, the frequency window spans the entire measured bandwidth ($N = \Nfrequency$). Consequently, only a single frequency-window position exists and the index $k_f$ can be omitted and the local scattering function for anchor $m$ at index time $k_t$, is estimated as
\begin{equation}\label{eq:local-scattering-functiom}
    \localScatteringMatrix{k_t} = \frac{1}{IJ}\sum_{w = 0}^{IJ-1} \left(\windowedH_{k_t} \odot \overline{\windowedH_{k_t}} \right).
\end{equation}
In \eqref{eq:local-scattering-functiom}, the matrix $\windowedH_{k_t}$ is the filtered channel matrix using the $I$ and $J$ separable band-limited discrete prolate spheroidal sequences: $I$ sequences in the frequency domain, and $J$ sequences in the time domain.
For details, see \cite{Nelson2024} where the relations between indices and matrices are explained thoroughly. 
For the analysis performed here, the parameters $I$ and $J$ were chosen as $I = 1$ sequence over frequency and $J = 2$ sequences over time \cite{Nelson2024}. 
The length $K$ of the window was first set to \num{150} to analyze the stationarity length in which the channel statistics is assumed to be wide-sense stationary. An overlap of the window of \SI{50}{\percent} was used. 

\begin{figure}[t]
	\centering
	\includegraphics{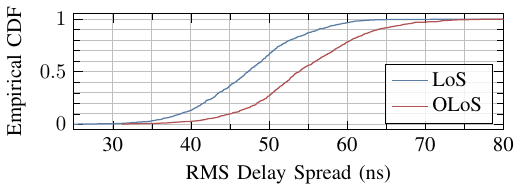}
	\caption{The \gls{rms} delay spread for the \gls{los} and \gls{olos} datasets.}
	\label{fig:rms-delay-spread}
\end{figure}

From the local scattering function it is straightforward to derive the \gls{pdp} and \gls{dsd} as the marginal expectations over the corresponding dimension. 
When calculating the \gls{rms} delay spread, the moments of the \gls{pdp} need to be calculated. 
To obtain accurate estimates of the \gls{rms} delay spread, only contributions from the \gls{pdp} that exceed certain power thresholds should be considered \cite{Czink2007PhD}. 
The thresholds were selected as \SI{5}{\dB} above the noise floor to mitigate spurious peaks, and \SI{40}{\dB} below the \gls{pdp} peak to only consider components with significant contributions. 
The \gls{ecdf} for the \gls{rms} delay spread is shown in Fig.~\ref{fig:rms-delay-spread} for the \gls{los} and \gls{olos} datasets, respectively. The delay spread is naturally smaller in \gls{los} than in \gls{olos}, with median values of 47 and 53~ns, respectively.

\subsection{Collinearity}\label{sec:collinear}
\begin{figure}[t]
\centering
\subfloat{\includegraphics{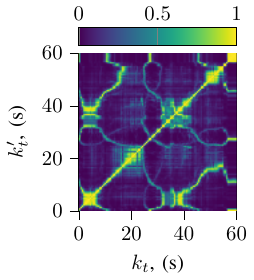}%
\label{fig:collinearity-ref3-a4}}
\hfil
\subfloat{\includegraphics{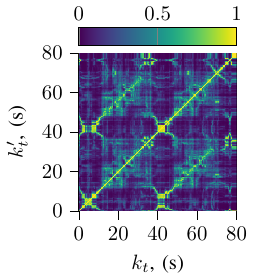}%
\label{fig:collinearity-loop4-a4}}
\caption{The collinearity matrix for two of the measured scenarios; ref (left) and loop (right). Both plots depict the collinearity between the agent and anchor \num{4}.}
\label{fig:collinearity}
\end{figure}

Using the collinearity metric between the local scattering functions in two different time instances $k_t$ and $k'_t$, results in a quantity of how similar the statistics are at the two different instances and, in extension, how long the stationarity region is. The collinearity metric $R\left(k_t, k'_t\right)$ is defined as

\begin{equation}
    R\left(k_t, k'_t\right) = \frac{\left(\vectorize\localScatteringMatrix{k_t}\right)^\trp  \left(\vectorize\localScatteringMatrix{k'_t}\right)}{\|\vectorize\localScatteringMatrix{k_t}\|\cdot\|\vectorize\localScatteringMatrix{k'_t}\|},
\end{equation}
where $\bm{C}$ represents the local scattering function in \eqref{eq:local-scattering-functiom}.
The collinearity between the agent and anchor \num{4} from the ref and loop scenario are shown in Fig.~\ref{fig:collinearity}. Anchor 4 is chosen for illustrative purpose, but the analysis holds for every anchor. All the elements of the resulting collinearity matrix are between zero and one (the diagonal is always one), and the matrix will always be symmetric. 
In Fig.~\ref{fig:collinearity-ref3-a4}, where the agent was driving back and forth in the middle of the hall, there is a symmetry for $k_t + k'_t > 55$ where the same pattern starts to appear again. This is, when the agent is turning and starts driving back to the start position, meaning that the agent will experience similar statistics on the path back. In Fig.~\ref{fig:collinearity-loop4-a4}, where the agent is driving two laps around machinery, it can also be clearly seen when the agent is close to the starting position after one lap and then how there is a parallel line approximately with a constant \SI{40}{\second} separation. 
A threshold needs to be set in order to decide how similar the statistics needs to be in order to be considered \gls{wssus}. Defining the indicator function as

\begin{figure}[t]
	\centering
	\includegraphics{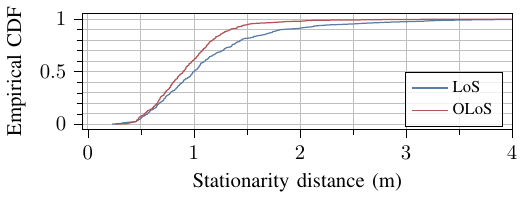}
	\caption{The \gls{ecdf}  of the estimated stationarity distance for the \gls{los} and \gls{olos} datasets, when the collinearity threshold was selected to be $c_\mathrm{th} = 0.9$.}
	\label{fig:stationarity-ecdf}
\end{figure}

\begin{equation}\label{eq:indicator-function}
\gamma\left(k_t, k'_t\right) =
    \begin{dcases*}
    1 & : $R\left(k_t, k'_t\right) > c_\mathrm{th}$\,, \\
    0 & : otherwise\,,
    \end{dcases*}
\end{equation}
where $c_\mathrm{th}$ defines the threshold. 
Following \cite{BernadoPhD}, the threshold was chosen as \num{0.9}.
Applying the indicator function \eqref{eq:indicator-function} to the collinearity results in a binary matrix with ones on the diagonal. The width of the region around the diagonal then gives the stationarity time.
By multiplying the stationarity region (expressed in seconds) by the average velocity of the agent, the stationarity distance can be acquired.
That is, the distance for which the \gls{wssus} assumption (approximately) holds. 
The \gls{ecdf} of the stationarity distance is plotted in Fig.~\ref{fig:stationarity-ecdf} for the \gls{los} and \gls{olos} datasets. 
It is clear that the median distance is on the order of \SI{1}{\metre} for the \gls{los} dataset, and slightly shorter for the \gls{olos} dataset. 
Based on these stationarity distances, a window length of \num{300} snapshots (\SI{1.5}{\second}) is selected when extracting the statistics in the remainder. Since the maximum instantaneous velocity of the agent was about \SI{0.8}{\metre\per\second}, the maximum length traveled during \num{300} snapshots approximately equals \SI{1.6}{\metre}, but is usually somewhat shorter.

\subsection{Small-scale fading statistics}\label{sec:smallScale}
\begin{figure}[t]
\centering
\includegraphics{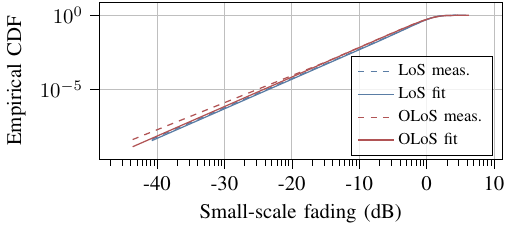}
\caption{The \glspl{ecdf} of the small-scale fading of the \gls{los} and \gls{olos} datasets, respectively, and their Ricean fitted distributions.}
\label{fig:ssf-cdfs}
\end{figure}

Let $\bm{A}^{(m)} = |\channelMatrix|$ be the instantaneous amplitude of the received signals. To obtain an estimate of the small-scale averaged amplitude, $\bm{A}^{(m)}_\text{SSA}\in \realset{\Nfrequency}{\Ntime - K + 1}$, a moving average window of length $K$ samples was applied over the time dimension for all tones. The window length $K$ should be chosen so that the channel is wide-sense stationary, which is usually set to approximately $10\lambda$ long~\cite{Molisch2012}, but due to the dynamic nature of the measured scenario, the value of $K$ will change over time. Hence, following the investigation of the stationarity region in Sec.~\ref{sec:collinear}, $K$ was set to 300 which will be approximately $10\lambda$ when the robot is moving at a constant velocity of \SI{0.8}{\metre\per\second}.
The estimate of the small-scale fading, $\bm{A}^{(m)}_\text{SSF}$, is then given as 
\begin{equation}\label{eq:ssf}
    \bm{A}^{(m)}_\text{SSF} = \left[\bm{A}^{(m)}\right]_{1:\Nfrequency,K/2:\Ntime-K/2 + 1} \odot \frac{1}{\bm{A}^{(m)}_\text{SSA}}, \in \realset{\Nfrequency}{\Ntime - K + 1},
\end{equation}
where $1/\bm{A}^{(m)}_\text{SSA}$ is the element-wise inversion under the assumption that none of the elements in $\bm{A}^{(m)}_\text{SSA}$ are zero. 
The \gls{ecdf}  of the small-scale fading, for the two datasets, is shown in Fig.~\ref{fig:ssf-cdfs} along with its Ricean fit that follows the measured curves. It is noteworthy that the small-scale fading shows similar behavior in the two datasets. 
The small-scale fading in the \gls{los} dataset can be modeled as $\Rice{0.84}{0.49}$ giving a $K$-factor of \num{1.44}.
For the \gls{olos} case, the distribution is modeled as a $\Rice{0.72}{0.59}$ distribution with a $K$-factor of \num{0.74}. The reason for the low K-factor in the \gls{los} dataset as well as the fact that there is a (although small) K-factor in the \gls{olos} dataset could be consequences of the classification of obstruction. It could also be attributed to the short link distances and rich scattering environment that could result in several strong multipath components.

\subsection{Channel hardening}\label{sec:hardening}
One key advantage of massive \gls{mimo} systems is the channel hardening effect. 
In essence, it means that massive \gls{mimo} can combat small-scale fading, mitigating fading dips due to destructive superposition of multipath components.
When the number of antennas $M$ increases, linear schemes such as \gls{mrt} have been shown to become optimal~\cite{MRT1999}. 
This effect can be very prominent in massive MIMO channels in rich scattering environments~\cite{Willhammar2024}. The system here cannot be considered massive MIMO, but rather consists of twelve distributed antennas; it is still shown in the following that there is a significant channel hardening effect.

In \cite{Willhammar2024} channel hardening for a distributed MIMO setup was measured in a limited area in a fully \gls{olos} scenario where channel parameters such as large-scale fading can be assumed to be constant. In addition, the distribution of the $M$ antennas is on a wall roughly \qtyproduct{10x3}{\metre} (W x H), which will not drastically influence the distance-dependent path gain.
In that setting, the complex channel gains can be modeled as independent and identically distributed (i.i.d.) complex Gaussian with no dominant component, which leads to a Rayleigh distribution of the channel gain amplitudes, and in extension, the channel power gains become exponentially distributed.
Furthermore, performing \gls{mrt} -- which translates to the sum of channel power gains with some normalization -- will lead to a Gamma distribution $\Gamma\left(M, 1/M\right)$ where $M$ is the number of antennas. 

\begin{figure}[t]
	\centering
	\includegraphics{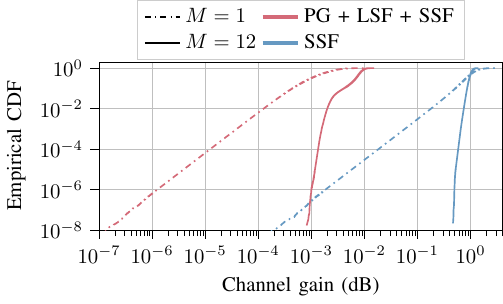}
	\caption{The \gls{ecdf} for the channel gain. Showing the effect with and without beamforming gain using \eqref{eq:mrt}. The red lines show the data under the influence of all propagation effects, i.e. path gain (PG), large-scale fading (LSF), and small-scale fading (SSF). In the blue lines the path gain and large-scale fading effects have been averaged out.}
	\label{fig:channel-hardening-cdf}
\end{figure}

In the datasets presented here, in the presence of time-varying distance-dependent path gain and large-scale fading, the channel hardening effect over time does not follow this theoretical distribution, in line with the analysis in~\cite{Willhammar2020}. 
Here, the small-scale fading distribution over frequency is inspected by performing \gls{mrt}.
For each time instant in the measurement set, i.e. $\forall k \in \{0, \ldots, \Ntime-1\}$, the following matrix~\cite{MRT1999, Willhammar2024}~is constructed
\begin{equation}
    \bm{H}_k = \left[\channelVecNoisy{1}{k}, \ldots, \channelVecNoisy{M}{k}\right] \in \complexset{\Nfrequency}{M},
\end{equation}
then the \gls{mrt} is computed as
\begin{equation}\label{eq:mrt}
    \bm{h}_k^{\mathrm{MRT}} = \left(\bm{H}_k^\herm \odot \bm{H}_k\right)\cdot\bm{1} \in \complexset{\Nfrequency}{1},
\end{equation}
without any normalization. In Fig.~\ref{fig:channel-hardening-cdf} the \gls{ecdf}  of the channel gain is plotted in log-log scale, with and without normalization such that the red curves show the channel gain including path gain, large-scale fading, and small-scale fading and the blue curves only show the influence of small-scale fading, both for one and twelve antennas when using MRT. Here, it can be seen that the deep dips are mitigated when combining the antennas, resulting in a steeper curve.
Due to the dynamic nature of the dataset there will be a varying mean power gain, but the deep dips have been mitigated.
Without normalization, large-scale fading effects are part of the results, giving a required fading margin of not much more than 10~dB. Considering small-scale fading only, just a few dB of fading margin is sufficient for reliable communication. In either case, one can conclude that D-MIMO is indeed an enabler of ultra-reliable communication and could contribute to reduced fading margins and reduce the need for retransmissions, also leading to reduced latencies.

\section{A distributed MIMO channel model for industrial environments}\label{sec:channelModel}

\subsection{Spatial consistency using Gaussian random fields}
\begin{figure}[t]
	\centering
	\includegraphics{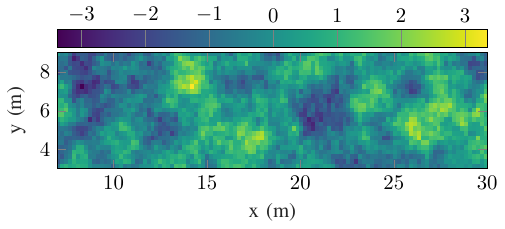}
	\caption{An example realization of a Gaussian random field (GRF) for the large-scale fading parameters, for one of the infrastructure anchors.}
	\label{fig:grf-lsp}
\end{figure}

We model the large-scale fading parameters with pre-computed two-dimensional \glspl{grf} that cover the simulation environment, one field per anchor, using the same measured exponential auto-correlation kernel. 
This is conceptually aligned with the spatial consistency procedure in 3GPP TR 38.901~\cite{3gppCHANNELMODEL}, where spatially correlated random variables are generated on a two-dimensional grid to get the prescribed correlation function. 
A comprehensive treatment of such grid-based spatial consistency for geometry-based stochastic channel models is also provided in \cite{ademaj2019spatial}, where it is shown that this approach preserves the desired marginal distributions and auto-correlation properties while ensuring position-dependent reproducibility. 
For computational efficiency, exact realizations of the \gls{grf} can be generated by circulant embedding of the covariance matrix \cite{dietrich1997fast}, which requires only a two-dimensional FFT over the grid. This approach achieves $O(N \log N)$ complexity, well suited for the indoor (i.e. bounded) environments considered here. As an alternative, the sum-of-sinusoids approach of \cite{jaeckel2018efficient} approximates the same spatial process with low memory requirements and has been adopted in \cite{3gppCHANNELMODEL}. 
Inter-anchor correlation is introduced by applying a Cholesky decomposition of the measured covariance matrix to the independently generated per-anchor fields, preserving both the spatial auto-correlation and the inter-anchor cross-correlation structure. An example realization of a \gls{grf} for one antenna is shown in Fig.~\ref{fig:grf-lsp}.

\begin{figure}[ht]
	\centering
	\includegraphics{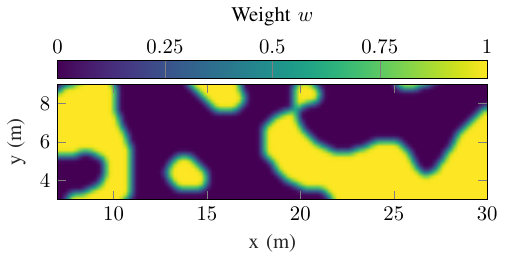}
	\caption{An example realization of a Gaussian random field (GRF) which is translated, based on a threshold, to create the binary \gls{los}/\gls{olos} state field for one anchor. The binary field is then convolved, in two dimensions, with a kernel to create the smooth transitions between states. A weight $w=1$ corresponds to \gls{los} and $w=0$ corresponds to \gls{olos}.}
	\label{fig:grf-state}
\end{figure}

A similar \gls{grf} technique is applied to the \gls{los}/\gls{olos} state process by thresholding a smooth spatial field with a calibrated threshold, yielding spatially consistent state transitions.
This thresholded \gls{grf} is the operative mechanism that determines the \gls{los}/\gls{olos} state in the simulator; the state-transition graph and rates in Sec.~\ref{sec:losProbability} are descriptive, characterizing the measured statistics to which this generative model is calibrated.
Unlike the large-scale fading fields, the state \gls{grf} uses a squared exponential kernel $R(d) = \exp\left(-d^2/d_{1/e}^2\right)$, to fulfill the requirements on differentiability for the level-crossing rate formula by Rice~\cite{Rice1944, jaeckel2018efficient}.
The threshold $\gamma_m = \Phi^{-1}(1 - p_{\text{LoS},m})$ is calibrated so that the fraction of space in \gls{los} matches the measured \gls{los} probability $p_{\text{LoS},m} = \bar{L}_{\text{LoS}} / (\bar{L}_{\text{LoS}} + \bar{L}_{\text{OLoS}})$, derived from the measured mean segment lengths. The decorrelation distance of the state \gls{grf} is then determined via Rice's formula for level crossings of Gaussian processes, ensuring that the mean segment lengths match the measured values. One example of a smoothed state transition field is shown in Fig.~\ref{fig:grf-state}.
This makes the model fully position-indexed and spatially consistent. Both large-scale fading and propagation states become deterministic functions of location, enabling reproducible channel realizations upon revisit while retaining the measured statistical properties.

\subsection{Soft state blending}
In transition regions between \gls{los} and \gls{olos}, the large-scale fading parameters are linearly interpolated based on a spatially consistent soft state weight $w_m(\mathbf{p}) \in [0,1]$, obtained by convolving the binary state field for anchor $m$ with a radially symmetric smoothing kernel:
\begin{equation}
    w_m(\mathbf{p}) = (b_m * h)(\mathbf{p}),
\end{equation}
where $b_m(\mathbf{p}) \in \{0,1\}$ is the thresholded state field and $h(r)$ is a normalized Hann kernel. The width of the kernel is set equal to the measured decorrelation distance (Sec.~\ref{sec:lspAutocorrelation}), so that the physical size of the transition region is tied to the distance over which the channel statistics remain correlated. The resulting large-scale fading is 
\begin{equation}
    \text{LSF}(\mathbf{p}) = \mu(w) + \sigma(w) \cdot G(\mathbf{p})
\end{equation}
where $\mu(w) = w \mu_{\text{LoS}} + (1-w) \mu_{\text{OLoS}}$ and $\sigma(w) = w \sigma_{\text{LoS}} + (1-w) \sigma_{\text{OLoS}}$. Since both state-dependent large-scale fading values derive from the same underlying \gls{grf} $G(\mathbf{p})$, this blending preserves Gaussianity of the marginal distribution while avoiding non-physical discontinuities at state transitions. Because $w_m(\mathbf{p})$ is purely a function of spatial location, and not the trajectory, the spatial correlation structure of the \gls{grf} is fully preserved without trajectory-dependent artifacts.

\subsection{Inter-anchor correlation model}
The inter-anchor correlation matrix models the cross-correlation of large-scale fading between different anchor links. Following the spatial correlation framework established in~\cite{Gudmundson1991}, we employ an exponential correlation kernel based on inter-anchor distance:
\begin{equation}
    \rho_{ij} = \exp\left(-\frac{d_{ij}}{d_{\text{corr}}^{\text{anchor}}}\right) + \epsilon
\end{equation}
where $d_{ij}$ is the distance between anchors $i$ and $j$, $d_{\text{corr}}^{\text{anchor}}$ is the decorrelation distance, and $\epsilon$ represents environment variations. The decorrelation distance is chosen to \SI{10}{\metre}, which is the correlation distance for the majority of parameters in \cite{3gppCHANNELMODEL} for the indoor factory scenario. The use of a noise term is motivated by measurements in \cite{weitzen2002measurement}, which found large cross-correlation variations, indicating significant variability around a purely deterministic model. Inspired by \cite{weitzen2002measurement}, we model the covariance with a distance-dependent cross-covariance matrix with an added noise term of $\epsilon = 0.3$, which is based on the measurements shown in Fig.~\ref{fig:lsf-cov}, but tweaked empirically due to the limited number of data points. To ensure that the covariance matrix is positive definite, the resulting matrix is projected to the nearest positive definite matrix~\cite{higham1988computing, szyszkowicz2010feasibility}.
This projection changes the original correlation values only marginally; in our empirical tests the change is smaller than the variance of the random realization drawn prior to projection, so it does not significantly alter the measured distance-dependent correlation. The values $d_{\text{corr}}^{\text{anchor}} = \SI{10}{\metre}$ and $\epsilon = 0.3$ should be regarded as defaults: the former follows the 3GPP indoor factory model and the latter is set from the measured spread, and both should be recalibrated for environments that differ substantially in size, clutter, or anchor layout from the one measured here.

\subsection{Channel model workflow}
In the following, all items for the channel model are summarized and a recipe for generating the channel model is provided. First, Table~\ref{tab:summary} summarizes the items that need to be set when generating channels and the items that have been estimated and modeled. For each item, the values and a brief description are included. Because the model is a geometry-based stochastic channel model,
the portability of the model parameters varies with how much a new environment differs from the one measured here. Parameters that are tied to the specific geometry and clutter of the hall should be refit for environments that differ substantially in layout or clutter. Structural choices that follow the 3GPP indoor factory framework
are expected to transfer to similar industrial halls without refitting.
\begin{table}[th]
    \centering
    \caption{Summary of the items in the channel model.}
    \begin{tabularx}{\linewidth}{l|X}
        \toprule
         Anchors & User-defined number of anchors and placements.\\
         \hline
         Agent & User-defined trajectory of the agent. \\
         \hline
         IOs & User-defined number of IOs. Stochastic placement of the IOs based on estimated \gls{rms} delay spread. \\
         \hline
         State transition & Stochastic based on the average distance in a state;  \gls{los}: \SI{4.65}{\metre}, \gls{olos}: \SI{5.55}{\metre}.\\
         \hline
         Path gain & Deterministic based on estimated parameters; \gls{los}: $k=-0.86$, $m=-44.24$ and \gls{olos}:  $k=-0.95$ and $m=-48.78$. \\
         \hline
         Large-scale fading & Stochastic based on estimated parameters; log-normal with variances $2.13$\,dB in \gls{los} and $3.25$\,dB in \gls{olos}. \\
         \hline
         Covariance & Stochastic based on inter-anchor distances and estimated parameter $\sigma = 0.28$\\
         \hline
         Auto-correlation & Deterministic based on estimated parameters; \gls{los}: $k=-0.82$ and \gls{olos}: $k=-0.81$. \\
         \hline
         Small-scale fading & Stochastic based on estimated parameters; \gls{los}: Rice(0.84, 0.49) and \gls{olos}: Rice(0.72, 0.59).\\  
         \hline
         Doppler spread & Stochastic and dependent on the placement of IOs and the user-defined trajectory. \\
         \bottomrule
    \end{tabularx}
    \label{tab:summary}
\end{table}

With all the parameters in place, here is a summary of how to apply these to generate realizations of the spatially consistent stochastic channel model, applicable for D-MIMO in industrial environments. The spatial consistency is achieved through GRFs, which ensure that large-scale fading and link-state parameters exhibit realistic spatial correlation properties. This means revisiting the same location produces identical channel realizations.              
The first step is the initialization phase:
\begin{enumerate}[A)]
      \item Select the chosen number of anchors $M$ and place them in the site-specific environment considered.
      \item Place the agent at its initial position in the environment.               \item Build the inter-anchor covariance matrix $\mathbf{C} \in \realset{M}{M}$; the diagonal elements are all ones, and the off-diagonal elements are determined by a distance-based spatial correlation model (Sec.~\ref{sec:lsp-covariance}).
      \item For each anchor $m$, generate an independent \gls{grf} $G_m(\mathbf{p})$ over the environment with an exponential kernel parameterized by the decorrelation distance (Sec.~\ref{sec:lspAutocorrelation}). These fields will be used for large-scale fading.
      \item Apply cross-anchor correlation to the large-scale fading fields via Cholesky decomposition: $\mathbf{G}_\mathrm{corr} = \mathbf{L}\,\mathbf{G}_\mathrm{indep}$, where $\mathbf{L}$ is the lower Cholesky factor of $\mathbf{C}$.
      \item For each anchor $m$, generate an independent state \gls{grf} $S_m(\mathbf{p})$ with a decorrelation distance calibrated from measured \gls{los}/\gls{olos} segment lengths (Sec.~\ref{sec:losProbability}).
      \item Compute the state threshold $\gamma_m$ for each anchor from the target \gls{los} probability: $\gamma_m = \Phi^{-1}(1 - P_{\mathrm{LoS},m})$, where $\Phi^{-1}$ is the inverse standard normal \gls{cdf}.                           
      \item Threshold each state \gls{grf} to obtain a binary field $b_m(\mathbf{p}) = \mathbf{1} [S_m(\mathbf{p}) > \gamma_m]$ and convolve it with a radially symmetric smoothing kernel to produce the soft state weight field $w_m(\mathbf{p}) \in [0,\,1]$ as described in Sec.~VII.B.
      \item Sample all \glspl{grf} along the agent trajectory $\mathbf{p}(k)$ and determine the link state for each anchor: \gls{los} if $S_m(\mathbf{p}(k)) > \gamma_m$, \gls{olos} otherwise.
      \item Generate and distribute IOs , in total $N_\mathrm{IO}$, for each anchor in the environment and draw an additional delay for each of them such that the power delay matches an exponential decay with an \gls{rms} delay spread corresponding to Fig.~\ref{fig:rms-delay-spread}. 
      These IOs are then used for simulation of the small-scale fading and represent the last interaction point seen from the agent. 
      The number of IOs, $N_\mathrm{IO}$, depends on the resolution of the antennas and should be sufficiently large to accurately model the small-scale fading, meaning that one should not be able to resolve individual MPCs in a cluster~\cite{Hoeher1992, 3gppCHANNELMODEL}.
    \item Initialize the random start phases of the anchors.
\end{enumerate}
After initialization, the next steps are the channel simulations for each anchor $m$ at time instance $k$:
  \begin{enumerate}
      \item Update the agent position $\mathbf{p}(k)$ according to its trajectory.
      \item Calculate the distances between the anchor, the agent, and the interacting objects.
      \item Read the soft state weight $w_m(k) \in [0,\,1]$ from the pre-computed smoothed state field (see initialization step~8), enabling gradual, spatially consistent transitions between \gls{los} and \gls{olos}.
      \item Calculate the deterministic path gain by blending between the \gls{los} and \gls{olos} path gain models: $G(k) = w_m(k)\, G_\mathrm{LoS} + (1 - w_m(k))\, G_\mathrm{OLoS}$; see Sec.~\ref{sec:pathGain}.                            
      \item Read the large-scale fading from the pre-sampled GRF and blend between states: $\ell_m(k) = w_m(k)\, \ell_{\mathrm{LoS},m}(k) + (1 - w_m(k))\, \ell_{\mathrm{OLoS},m}(k)$. The temporal correlation arises implicitly from the      
  spatial correlation of the GRF as the agent traverses the field (Sec.~\ref{sec:lsp}). The instantaneous $K$-factor is blended analogously.                                                 \item Simulate small-scale fading according to Sec.~\ref{sec:smallScale}, using the blended $K$-factor.                                                                       \item Compose the channel transfer function: path gain $\times$ large-scale fading $\times$ small-scale fading.
\end{enumerate}

The model is inspired by the COST 2100 framework, both being geometry-based stochastic channel models that, based on scatterers in the environment, can simulate MIMO channels over time, frequency and space~\cite{COST2100}. This approach is here extended to a D-MIMO scenario, meaning that correlations of large-scale fading parameters are important to capture in the model. The scenarios in the COST 2100 framework capture typical indoor and outdoor environments, while here the extraction of channel characteristics for an industrial scenario is the focus. 
The parameters are derived from the measured scenario with distributed single antennas in the environment. Future possible extensions are the inclusion of clusters and visibility regions of clusters and scatterers to account for antenna correlations in cases where the agent or anchors have multiple antennas. The current setup does not allow for such analysis. In the case where many antennas are used at the anchors or agent, or if a wider system bandwidth is used, the number of scatterers has to increase to match the resolvability of the system.

\begin{figure}[th!]
	\centering
	\includegraphics{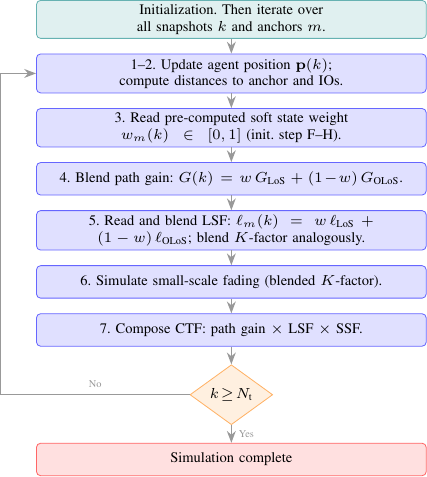}
	\caption{Flowchart of how to generate a realization of the distributed MIMO channel model for industrial environments.}
	\label{fig:flow}
\end{figure}

\section{Validation of the model}\label{sec:validation}
\begin{figure}[t]
	\centering
	\includegraphics{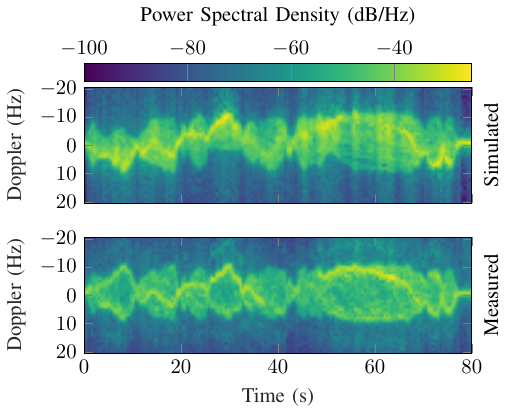}
	\caption{Comparing simulation (top) against measurements (bottom). This is the Doppler spread in a section of the scan scenario. It is clear that the essential characteristics of the Doppler spectrum are captured in the simulation.}
	\label{fig:valid-meas-vs-sim}
\end{figure}

\begin{figure}[t]
	\centering
	\includegraphics{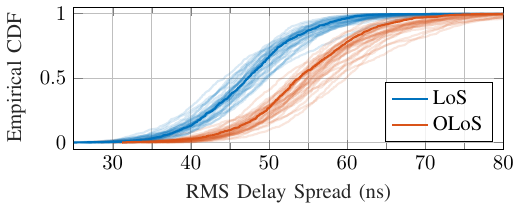}
	\caption{The simulated \gls{rms} delay spread when portions of the scan and ref scenarios (the \includegraphics{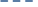} and \includegraphics{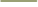} trajectories in Fig.~\ref{fig:scenarios}) have been simulated. The two thicker lines are the measured \gls{rms} delay spreads, as shown in Fig.~\ref{fig:rms-delay-spread}.}
	\label{fig:valid-sim-rms-delay}
\end{figure}

The proposed simulation framework has been implemented for validation using the recorded ground truth trajectory from the \emph{scan} scenario and the recorded locations of the anchors as inputs. The \emph{scan} scenario can be considered representative of the measured environment, as it covers most parts of the hall and includes both \gls{olos} and \gls{los} conditions. The system parameters are the same as during the measurement campaign; i.e., a carrier frequency of \SI{3.75}{\giga\hertz} using \num{449} carriers separated by \SI{78.125}{\kilo\hertz} resulting in a bandwidth of \SI{35}{\mega\hertz}. The implementation is open source and can be found online \cite{Nelson2025zenodo}. In Fig.~\ref{fig:valid-meas-vs-sim} the Doppler profiles for both simulation and measurement are presented using $N_\mathrm{IO} = 50$ interacting objects per anchor. It is clear that the simulator accurately captures the \gls{los} component and the essential characteristics of the Doppler spectrum. 
The variation of the simulated Doppler profiles across realizations is expected and physically meaningful. As a geometry-based stochastic model, each realization places the IOs at different positions drawn from the same statistics, so an individual profile reflects one valid scatterer geometry while the collection reproduces the measured behavior on average. 
To further validate the model, Fig.~\ref{fig:valid-sim-rms-delay} shows the \gls{rms} delay spreads when parts of the scan and ref scenarios are simulated for 100~realizations and divided into the \gls{los} and \gls{olos} states. The thicker lines are the measured \gls{rms} delay spreads from Fig.~\ref{fig:rms-delay-spread}. The simulated realizations show, on average, a good agreement with the measurements.
Although only the Doppler spectra and \gls{rms} delay spread are shown here, the received-power statistics are reproduced by construction, since the path gain and the large-scale fading are extracted directly from the measured received power; moreover, as the Doppler and delay characteristics depend on the same propagation geometry, their agreement with the measurements would be unlikely if the power levels were not also well represented.
A direct validation of the inter-anchor covariance against simulated realizations is constrained by the limited measured data, which provide only a single spatial realization of the covariance. The reported covariance therefore reflects the best estimate obtainable from the available measurements.
Supporting the applicability of the inter-anchor model, the measurements show a $\sigma\approx0.28$ for the off-diagonal elements, where the corresponding value for the model ($\sigma\approx0.29$) shows good agreement.

\section{Conclusions}\label{sec:conclusions}
A comprehensive analysis has been conducted based on data collected from a unique D-MIMO channel measurement campaign in an industrial environment. A new approach for classifying obstruction that is derived from lidar data and approximated from the first Fresnel zone is presented, including a state transition graph. The assumption that there are almost always (a few or several) strong links in D-MIMO systems is confirmed, and a quantification based on measurement data is provided, showing what one can expect from a real scenario. 
Another characteristic of D-MIMO that supports ultra-reliable communication is the increased potential for experiencing independent channels at the different antennas. The results show that there is not only a clear channel hardening effect reducing the small-scale fading effects but also, by evaluating the covariance matrix, it is established that the large-scale fading characteristics also show a diversity gain; an effect that is more prominent in \gls{los} than in \gls{olos}. In our scenario a 10~dB fading margin is sufficient for URLLC with negligible outage when using MRT.
In dynamic D-MIMO scenarios there are both spatial and temporal non-stationarities that need to be taken into account, which are here thoroughly investigated.
Finally, key channel parameters such as path gain, large-scale fading, small-scale fading, and \gls{rms} delay spread are evaluated; these are essential in order to achieve an accurate channel model, for which we here provide a step-by-step recipe to achieve spatial consistency, and that can be used for system development and evaluations of D-MIMO in similar industrial environments. For industrial environments with moving machinery or robots, further measurements are needed to capture the channel effects on Doppler and the parameters concerning correlation in time. The channel model has been implemented and validated, showing a good match and highlighting the applicability of the model.

\appendices
\section*{Acknowledgment}\label{sec:ack}
The authors extend their appreciation to Xuhong Li for her invaluable feedback during the data analysis. She also contributed to the measurement together with Aleksei Fedorov, Michiel Sandra, and Anders J. Johansson. The authors acknowledge the use of Claude Opus 4.6 for the code that simulates Gaussian random fields.

\ifCLASSOPTIONcaptionsoff
  \newpage
\fi

\bibliographystyle{IEEEtran}
\bibliography{bibtex/bib/IEEEabrv,bibtex/bib/ms}

\vfill

\end{document}